\documentclass[manuscript,preprint,onecolumn,aps,pre,eqsecnum,longbibliography,showpacs,showkeys,a4paper]{revtex4-1}
\usepackage[latin9]{inputenc}
\usepackage{color}
\definecolor{note_fontcolor}{rgb}{0.80078125, 0.80078125, 0.80078125}
\usepackage[english]{babel}
\usepackage{amsmath}
\usepackage{amssymb}
\usepackage{graphicx}
\usepackage{setspace}
\usepackage{esint}
\usepackage[unicode=true,
 bookmarks=true,bookmarksnumbered=true,bookmarksopen=true,bookmarksopenlevel=1,
 breaklinks=true,pdfborder={0 0 0},backref=false,colorlinks=true]
 {hyperref}
\hypersetup{
 pdfauthor={AINS}}

\makeatletter

\newenvironment{lyxgreyedout}
  {\textcolor{note_fontcolor}\bgroup\ignorespaces}
  {\ignorespacesafterend\egroup}

\@ifundefined{textcolor}{}
{%
 \definecolor{BLACK}{gray}{0}
 \definecolor{WHITE}{gray}{1}
 \definecolor{RED}{rgb}{1,0,0}
 \definecolor{GREEN}{rgb}{0,1,0}
 \definecolor{BLUE}{rgb}{0,0,1}
 \definecolor{CYAN}{cmyk}{1,0,0,0}
 \definecolor{MAGENTA}{cmyk}{0,1,0,0}
 \definecolor{YELLOW}{cmyk}{0,0,1,0}
 }
\numberwithin{equation}{section}
\numberwithin{figure}{section}
\numberwithin{table}{section}

\usepackage{graphicx}

\DeclareGraphicsExtensions{.png,.pdf,.eps}

\makeatother

\begin{document}

\title{A Classical Framework for Nonlocality and Entanglement\vspace*{\bigskipamount}
}

\author{Gerhard \surname{Grössing}\textsuperscript{1}}

\email[E-mail: ]{ains@chello.at}

\homepage[Visit: ]{http://www.nonlinearstudies.at/}

\selectlanguage{english}%

\author{Siegfried \surname{Fussy}\textsuperscript{1}}

\email[E-mail: ]{ains@chello.at}

\homepage[Visit: ]{http://www.nonlinearstudies.at/}

\selectlanguage{english}%

\author{Johannes \surname{Mesa Pascasio}\textsuperscript{1,2}}

\email[E-mail: ]{ains@chello.at}

\homepage[Visit: ]{http://www.nonlinearstudies.at/}

\selectlanguage{english}%

\author{Herbert \surname{Schwabl}\textsuperscript{1}}

\email[E-mail: ]{ains@chello.at}

\homepage[Visit: ]{http://www.nonlinearstudies.at/}

\selectlanguage{english}%

\affiliation{\textsuperscript{1}Austrian Institute for Nonlinear Studies, Akademiehof\\
 Friedrichstr.~10, 1010 Vienna, Austria}

\affiliation{\textsuperscript{2}Institute for Atomic and Subatomic Physics, Vienna
University of Technology\\
Operng.~9, 1040 Vienna, Austria\vspace*{2cm}
}

\date{\today}
\begin{abstract}
Based on our model of quantum systems as emerging from the coupled
dynamics between oscillating ``bouncers'' and the space-filling
zero-point field, a sub-quantum account of nonlocal correlations is
given. This is explicitly done for the example of the ``double two-slit''
variant of two-particle interferometry. However, it is also shown
that the entanglement in two-particle interferometry is only a natural
consequence of the fact that already a ``single'' two-slit experiment
can be described on a sub-quantum level with the aid of ``entangling
currents'' of a generally nonlocal nature.%
\begin{lyxgreyedout}
\global\long\def\VEC#1{\mathbf{#1}}

\global\long\def\d{\,\mathrm{d}}

\global\long\def\e{{\rm e}}

\global\long\def\meant#1{\left<#1\right>}

\global\long\def\meanx#1{\overline{#1}}

\global\long\def\mpbracket{\ensuremath{\genfrac{}{}{0pt}{1}{-}{\scriptstyle (\kern-1pt +\kern-1pt )}}}

\global\long\def\pmbracket{\ensuremath{\genfrac{}{}{0pt}{1}{+}{\scriptstyle (\kern-1pt -\kern-1pt )}}}

\global\long\def\p{\partial}
\end{lyxgreyedout}

\end{abstract}


\maketitle

\section{Introduction\label{sec:intro}}

Although nonlocality has featured very prominently throughout the
last decades in the discussions on the foundations of quantum mechanics,
no general consensus has yet been reached over it. Despite many arguments
in favor of the position that nonlocal effects are one of the (if
not\emph{ the}) main characteristics which distinguish classical from
quantum mechanics, some researchers even hold on to the view that
a purely local physics could suffice to explain all existing experimental
data, thus ultimately relying on a familiar trait of classical physics.

However, even if one considered quantum mechanical nonlocality as
a well-established fact, as we do, this does not necessarily mean
that there cannot exist some form of ``classical'' explanation for
it. In arguing for the use of modern, ``21\textsuperscript{st} century
classical physics'', our group, for example, has in recent years
obtained with such classical means a series of results that were previously
considered as obtainable only via quantum mechanics. Among these results,
some features figure prominently, like, e.g., explanations of Planck's
postulate of energy quantization, the dispersion of a Gaussian, or
interference at a double slit. (For an introduction, see \cite{Grossing.2012quantum}
and the references therein.) Thus, as a further task along our lines
of reasoning, we intend in this paper to provide a classical framework
for nonlocal effects and entanglement.

One of our main modeling scenarios is provided by the stimulating
experiments of Couder's group \cite{Couder.2005,Couder.2006single-particle,Protiere.2006,Eddi.2009,Fort.2010path-memory},
where ``bouncing'', and also ``walking'', droplets (the ``particles'')
are dynamically coupled to the oscillations of a bath (the ``waves'')
and thus produce a whole series of effects, which were previously
considered only to be possible quantum mechanically. Among these effects,
interference at a double slit, tunneling, or quantization of angular
momentum in closed orbits could be reproduced, for example. Although
it is clear that the mentioned experiments can only provide analogies,
at best, one has here, nevertheless, a scenario providing essential
stimuli for model building also in the context of quantum theory.
This, at least, is what we want to propose here, i.e., that there
are further insights to be gained from the experiments of Couder's
group, which could analogously be transferred into the modeling of
quantum behavior. Concretely, we do believe that also an understanding
of nonlocality and entanglement can profit from the study of said
experiments. In fact, one indispensable prerequisite for these experiments
to work, one basic commonality of all of them, is that the bath is
vibrating itself. It so happens that bouncer and bath may engage in
a self-organized, dynamically coupled entity where the bouncer can
self-propel due to its interaction with the wave it generates.

In recent papers, we introduced an analogous scenario in the quantum
domain, thereby having a particle/bouncer undergo also stochastic
jumps such that diffusion theory can be applied. Then, what is in
a quantum mechanical context described as a Gaussian wave-packet has
an equivalent in the Gaussian distribution of a particle whose path
follows the ``agitations'', or ``excitations'', of the underlying
``bath''. With the latter, we refer in our model to the zero-point
oscillations of the vacuum, i.e., something we take as empirically
given, or, in other words, as an ``ontological'' input to our theory.
Note that we completely agree here with Timothy Boyer who noted in
a similar context:\medskip{}

\begin{minipage}[t]{0.95\columnwidth}%
\textquotedblleft{}The concept of zero-point radiation (random radiation
fluctuations at the zero of temperature) can appear in both classical
and quantum theories. Zero-point radiation can not be regarded as
belonging exclusively to quantum theory any more than the concepts
of mass, energy, and gravity can be claimed as exclusively classical
concepts because they appeared first in the context of classical mechanics.\textquotedblright{}
\cite{Boyer.2010blackbody}%
\end{minipage}\medskip{}

As our model's particle through its bouncing and locking-in with the
zero-point field creates diffusion wave fields, and as the latter,
at least in the nonrelativistic case, extend instantaneously along
nonlocal distances \cite{Mandelis.2000diffusion,Mandelis.2001structure,Mandelis.2001diffusion-wave},
we speak of a nonlocal ``path excitation field''. Wherever the diffusion
wave fields ``radiated out'' into the environment exist (or better:
Wherever the bouncers' oscillations lock in with the oscillations
of the space-filling bath), the self-propelled bouncer (= ``walker'')
may go, eventually. So, in the case of a bouncer oscillating in some
source region of an interferometer experiment, whenever it is propelled
forward and can potentially go through one of the two slits, the (Gaussian)
path excitation fields behind the double-slit will overlap to produce
the familiar fringes at some screen \cite{Groessing.2012doubleslit}.
We stress, however, that one of the most important features of the
path excitation field is its nonlocal nature. This is actually what
we are going to work out in more detail in the present paper.

Apart from accepting as given the nonrelativistic diffusion wave fields'
``breathing'' nonlocally throughout space, what could be a possible
origin of the assumed nonlocal nature of the zero-point oscillations?
One can only speculate at present, but it seems that a good candidate
for an explanation would come from cosmological considerations. Note,
for example, that for the universe in its initial phases, according
to present-day models, one can admit, in addition to the particles
existing in the very early universe, a set of phase-locked wave-like
oscillations that would thus ``resonate'' throughout the whole small-scale
universe. Then, it is conceivable that cosmic inflation, for example,
would not destroy these oscillations, but rather ``inflate'' these
fields as well, thus ending up with a much larger universe where the
particles still oscillate in phase with the zero-point background,
albeit with the latter now having turned a nonlocal one. Whatever
the true reason for the nonlocality of the zero-point field may eventually
be, we take the latter as an input for our modeling, and in the following
show some results thereof.

\section{A Sub-Quantum Kinematic Account of Two-Particle Correlations }

\begin{figure}[!h]
\begin{centering}
\includegraphics{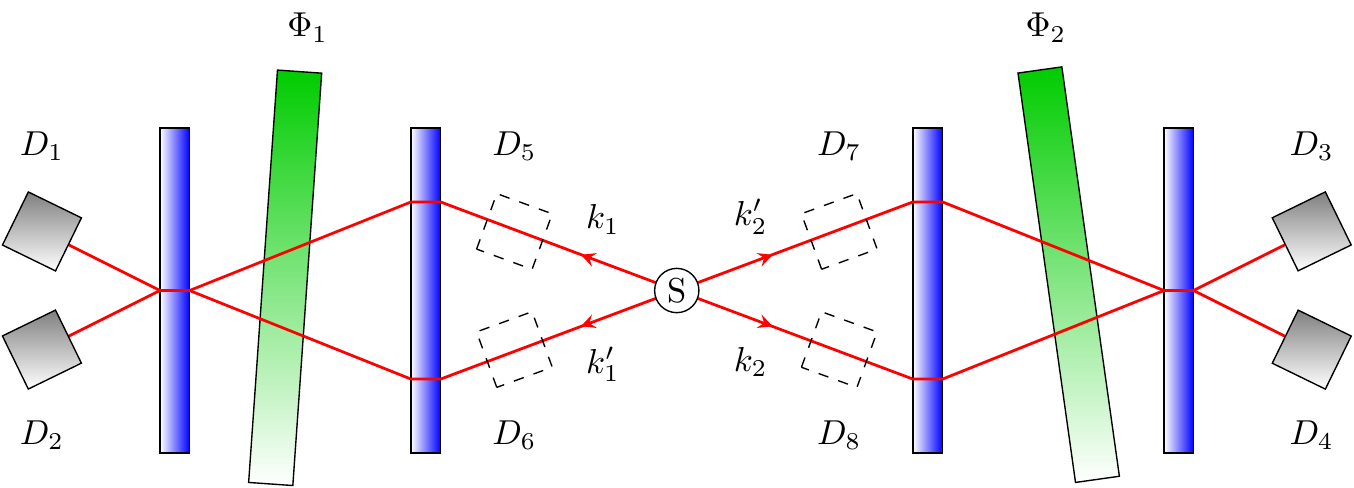}
\par\end{centering}

\caption{Scheme of a two-particle interferometer, with the source \emph{S}
in the center emitting two anti-correlated particles, with different
phase shifters $\Phi$ inserted into the particle beams, and with
(possible locations of) detectors \emph{D}.\label{fig1}}
\end{figure}

Let us first consider an EPR-type experiment as it was proposed by
Horne and Zeilinger~\cite{Horne.1986einstein-podolsky-rosen} for
two-particle interferometry, i.e., essentially a ``doubled'' double-slit
experiment as shown in Fig.~\ref{fig1}. One creates a particle pair
with equal but opposite momenta $\VEC k_{1}$ and $\VEC k_{2}$, respectively,
along with some small changes $\delta\VEC k$, which can be varied
via phase shifters $\Phi_{1}$ and $\Phi_{2}$. Their quantum mechanical
wave function would be described in terms of an entangled state. The
resulting correlated intensity $I\left(\VEC x_{1},\VEC x_{2}\right)$
exhibits marked modulations, which persist even for arbitrarily large
spatial separations of the individual particles' wave packets.

In our approach, we can make use of the path excitation field as follows.
Each path $i$ be occupied by a Gaussian wave packet with a ``forward''
momentum $\VEC p_{i}=\hbar\VEC k_{i}=m\VEC v_{i}$. (Moreover, due
to the stochastic process of path excitation, the latter is represented
also by a large number of consecutive Brownian shifts, $\VEC p_{u,\alpha}=m\VEC u_{\alpha}$,
but we shall return to this only in the next chapter, where we shall
discuss in more detail the different roles played by the velocity
fields $\VEC v$ and $\VEC u$, respectively.) To start with, we note
first the probability density currents for each ``particle'', and
then combine them in a suitable manner. Thus, upon preparation of
the two-particle source, we start with some initial distribution $P\left(\VEC x_{0},\, t_{0}\right)=R_{0}^{2}\left(\VEC x_{0},\, t_{0}\right)$
of a composite system just at decay time $t_{0}$. Considered as a
two-particle system, and since we employ a wave theory, this initial
distribution can be rewritten as $P\left(\VEC x_{0},\, t_{0}\right)=R_{L}\left(\VEC x_{0},\, t_{0}\right)R_{R}\left(\VEC x_{0},\, t_{0}\right)$,
with $R_{L}$ and $R_{R}$ signifying normalized amplitudes associated
with particles going left (locations $\VEC x_{1}$) or right (mirror
locations $\VEC x_{2}$), respectively. This initial distribution
is thus split up into one channel for particle \emph{$1$ }and a correspondingly
anti-correlated channel for particle $2$ (Fig.~\ref{fig1}). Note
that, because \textit{each of the two bouncers ``excites'' the areas
in both directions (i.e., both possible paths)} of the surrounding
medium, the probability density \textit{for each path} is given by
$P\left(\VEC x_{i},\, t\right)=R_{L}\left(\VEC x_{i},\, t\right)R_{R}\left(\VEC x_{i},\, t\right)$,
for $i=1$ or $2$, and similarly for the primed quantities, where
$R_{L}$ and $R_{R}$ now more specifically refer to the amplitudes
``excited'' by the bouncer going left or right, respectively, irrespective
of which actual particle one focuses upon. For simplicity, we shall
in the examples below concentrate on the symmetric scenario with equal
weights, i.e., $R_{L}\left(\VEC x_{1}\right)=R_{R}\left(\VEC x_{1}\right)=R_{L}\left(\VEC x_{2}\right)=R_{R}\left(\VEC x_{2}\right)$.

Then, in close similarity to our treatment of the case of a single
double-slit~\cite{Groessing.2012doubleslit}, for the ``doubled''
setup we write down the total average probability current as the sum
of all four average probability currents present, i.e., firstly without
the presence of any phase shifters it holds (with bars denoting averages)
that
\begin{equation}
\meanx{\VEC J}_{\textrm{tot}}=P_{{\rm tot}}\meanx{\VEC v}_{{\rm tot}}=P_{{\rm tot}}\frac{\hbar}{m}\meanx{\VEC k}_{{\rm tot}}:=\frac{\hbar}{m}\left[R_{L}R_{R}\VEC{\overline{k}}_{1}+R_{L}R_{R}\VEC{\overline{k}}_{2}+R_{L}'R_{R}'\VEC{\overline{k}}_{1}'+R_{L}'R_{R}'\VEC{\overline{k}}_{2}'\right].
\end{equation}
Allowing now for a relative phase $\varphi=\Phi_{1}-\Phi_{2}$ through
the insertion of the phase shifters $\Phi_{1}$ and $\Phi_{2}$, one
can write with the ``total'' average momenta $\VEC{\overline{k}}:=\VEC{\overline{k}}_{1}+\VEC{\overline{k}}_{2}$
and $\VEC{\overline{k}'}:=\VEC{\overline{k}}_{1}'+\VEC{\overline{k}}_{2}'+\overline{\VEC{\delta k}}$,
respectively,

\begin{equation}
P_{\mathrm{tot}}\meanx{\VEC k}_{\mathrm{tot}}=\left[R_{L}R_{R}\left(\VEC{\overline{k}}_{1}+\VEC{\overline{k}}_{2}\right)+R_{L}'R_{R}'\left(\VEC{\overline{k}}_{1}'+\VEC{\overline{k}}_{2}'+\overline{\VEC{\delta k}}\right)\right].
\end{equation}

Then we obtain with normalization $\mathfrak{\mathsf{\mathcal{N}}}$,
with the momentum balance $k_{\mathrm{tot}}=k=k'$, and with hats
denoting average unit vectors, the correlated intensity
\begin{equation}
I\left(\VEC x_{1},\VEC x_{2}\right):=\mathcal{N}^{2}P_{\mathrm{tot}}^{2}\left(\VEC x_{1},\VEC x_{2}\right)=\mathcal{N}^{2}\left[R_{L}R_{R}\VEC{\hat{k}}+R_{L}'R_{R}'\VEC{\hat{k}}'\right]^{2},
\end{equation}
and thus
\begin{align}
I\left(\VEC x_{1},\VEC x_{2}\right)= & \mathcal{N}^{2}\left[\vphantom{\int_{0}^{0}}R_{L}^{2}R_{R}^{2}+R_{L}'^{2}R_{R}'^{2}\right.\nonumber \\
 & +\left.2R_{L}R_{R}R_{L}'R_{R}'\cos\left\{ \left[\left(\VEC{\overline{k}}_{1}+\VEC{\overline{k}}_{2}\right)-\left(\VEC{\overline{k}}_{1}'+\VEC{\overline{k}}_{2}'+\overline{\VEC{\delta k}}\right)\right]\cdot\VEC r\right\} \vphantom{\int_{0}^{0}}\right].\label{eq:vax.4}
\end{align}
This is the exact quantum mechanical result, albeit here obtained
without invoking the quantum mechanical calculus. Moreover, one can
now also highlight particular features of this remarkable correlation
along nonlocal distances $r=x_{1}+x_{2}$ between the locations $\VEC x_{1}$
and $\VEC x_{2}$, respectively. Namely, as opposed to the formula~(\ref{eq:vax.4})
derived from the total probability density current, one can also single
out correlations between individual currents, respectively, for the
various channels. An important role is thereby played by the relative
phase $\varphi$, which emerges naturally in our model as it relates
a ``bouncer's'' oscillations along different paths \cite{Groessing.2012doubleslit,Groessing.2013dice}.

As is usual in interferometry, differences in relative phase are --
also classically -- accounted for by phase shifts of $\frac{\pi}{2}$
for each reflection of a beam at one of the slabs of the interferometer.
Thus, by comparing primed versus unprimed scenarios, one can for example
relate the intensities at detectors $D_{2}$ and $D_{4}$ in Fig.~\ref{fig1},
which provide the conditional probability

\begin{align}
P\left(D_{2}\left|D_{4}\right.\right)= & \mathcal{N}^{2}R_{L}^{2}R_{R}^{2}\left[\vphantom{\int_{0}^{0}}2+2\cos\left\{ \left[\left(\VEC{\overline{k}}_{1}+\VEC{\overline{k}}_{2}\right)-\left(\VEC{\overline{k}}_{1}'+\VEC{\overline{k}}_{2}'+\overline{\VEC{\delta k}}\right)\right]\cdot\VEC r\right\} \right]\nonumber \\
= & \frac{1}{4}\left(2+2\cos\left[\Phi_{1}-\Phi_{2}\right]\right)=\frac{1}{2}\left(1+\cos\varphi\right),\label{eq:vax.5}
\end{align}
whereas

\begin{equation}
P\left(D_{2}\left|D_{3}\right.\right)=\frac{1}{4}\left(2+2\cos\left[\Phi_{1}-\Phi_{2}+\pi\right]\right)=\frac{1}{2}\left(1-\cos\varphi\right).\label{eq:vax.6}
\end{equation}
However,

\begin{equation}
P\left(D_{6}\left|D_{4}\right.\right)=\frac{1}{4}\left(2+2\cos\frac{\pi}{2}\right)=\frac{1}{2},
\end{equation}
for example, because the phase difference between the two possible
paths is now independent of the $\Phi_{i}$ and given by $\varphi=\frac{\pi}{2}$.
From the latter example one sees that an ``early'' detection of
a particle at $D_{6}$, which is equivalent to a which-way measurement,
or to the ``closing of the second slit'' on one side of a double-double-slit
experiment, respectively, destroys nonlocal interference effects such
as those indicated by (\ref{eq:vax.5}) or (\ref{eq:vax.6}).

How can we understand this behavior in our model? Now, we have mentioned
on several occasions that we consider the Gaussians employed in our
calculations as simple solutions of a diffusion equation. More generally,
however, as has been explicated in the works of Mandelis et~al.~\cite{Mandelis.2000diffusion,Mandelis.2001structure,Mandelis.2001diffusion-wave},
diffusion wave fields related to oscillating sources may in a more
specific way extend nonlocally across the whole domain of an experimental
setup, for example. Thus, the Gaussians used so far may be only approximations
to more complex solutions of the diffusion equation, which actually
may exhibit long wiggly tails, albeit with very small amplitudes.
However, it is exactly such a functional characteristic which has
been found also in a specific quantum mechanical context, i.e., for
example, in Rauch's post-selection experiments \cite{Rauch.2006hidden,Rauch.2012particle}.
There, it turned out that for interference to occur it is not necessary
that the ``main bulks'' (approximated by Gaussians) of wave-packets
overlap. Rather, the experiments show that interference can be caused
by the nonlocally far-reaching action of the plane-waves of a quantum
mechanical wave-function. This is actually the corollary of our understanding
of the nonlocal nature of the zero-point field to which our particle/oscillator
couples: A Gaussian indicating the approximate whereabouts of our
bouncer, embedded in an oscillatory ``bath'' of momentum fluctuations,
the regular part of which thereby coinciding with the action of the
plane-wave components in quantum mechanics.

Note particularly that the bandwidth of these plane-wave components
is determined by the momentum resolution of the whole measurement
apparatus, where the upper limit is defined by the inverse of the
distance between source and detector \cite{Rauch.2012particle}. This
is strongly reminiscent of our analogy with the particle's bounces
in the Couder experiments, which are locked-in with the oscillations
of the fluid. The spatial constraints of the latter, i.e., the container
sizes, thus define the momentum resolution of the experiment via a
suitable bandwidth of possible wavelengths. So, we observe that in
our model it is the wholeness of the possible wave configurations
within the possibly nonlocal limits of an experimental setup that
co-determines the experiments' outcomes.

In other words, whenever the ``constraints'' of the experimental
setup are changed, this may have a nonlocal effect on the registered
particles -- a possibility which brings us back to the effect of the
``closing of a slit'' in the double-slit experiment. Note that,
as this effect is shown in a forthcoming paper~\cite{Groessing.2013dice}
to be essentially nonlocal, it is of the same nature as the one in
EPR-type experiments, as, for example, in Aspect's experiments: As
opposed to mere \emph{kinematic nonlocality} like the one implied
by intensity correlations such as~(\ref{eq:vax.4}), we are in ref.\textcolor{red}{~}\cite{Groessing.2013dice}
interested in the effects of the actual ``closing of a slit'', i.e.,
in \emph{dynamic nonlocality} \cite{Tollaksen.2010quantum}. However,
in the present paper we shall restrict our discussion to the simple
double-slit experiment. For this alone already suffices to bring forth
essential features of nonlocality and entanglement: The classical
roots of entanglement are already visible in the ``single'' double-slit
experiment, as will be shown now.

\section{Entangling Currents in the Double-Slit Experiment}

We consider the particles as emerging from one of two ``Gaussian
slits'', i.e., two possible paths of a particle which later eventually
cross each other. To do so, we started in \cite{Groessing.2012doubleslit}
with an independent numerical computation of two Gaussian wave packets,
with the total distribution given by
\begin{equation}
P_{{\rm tot}}:=P_{1}+P_{2}+2\sqrt{P_{1}P_{2}}\cos\varphi_{12}\,,\label{eq:vax.8}
\end{equation}
\begin{figure}[p]
\centering{}\includegraphics{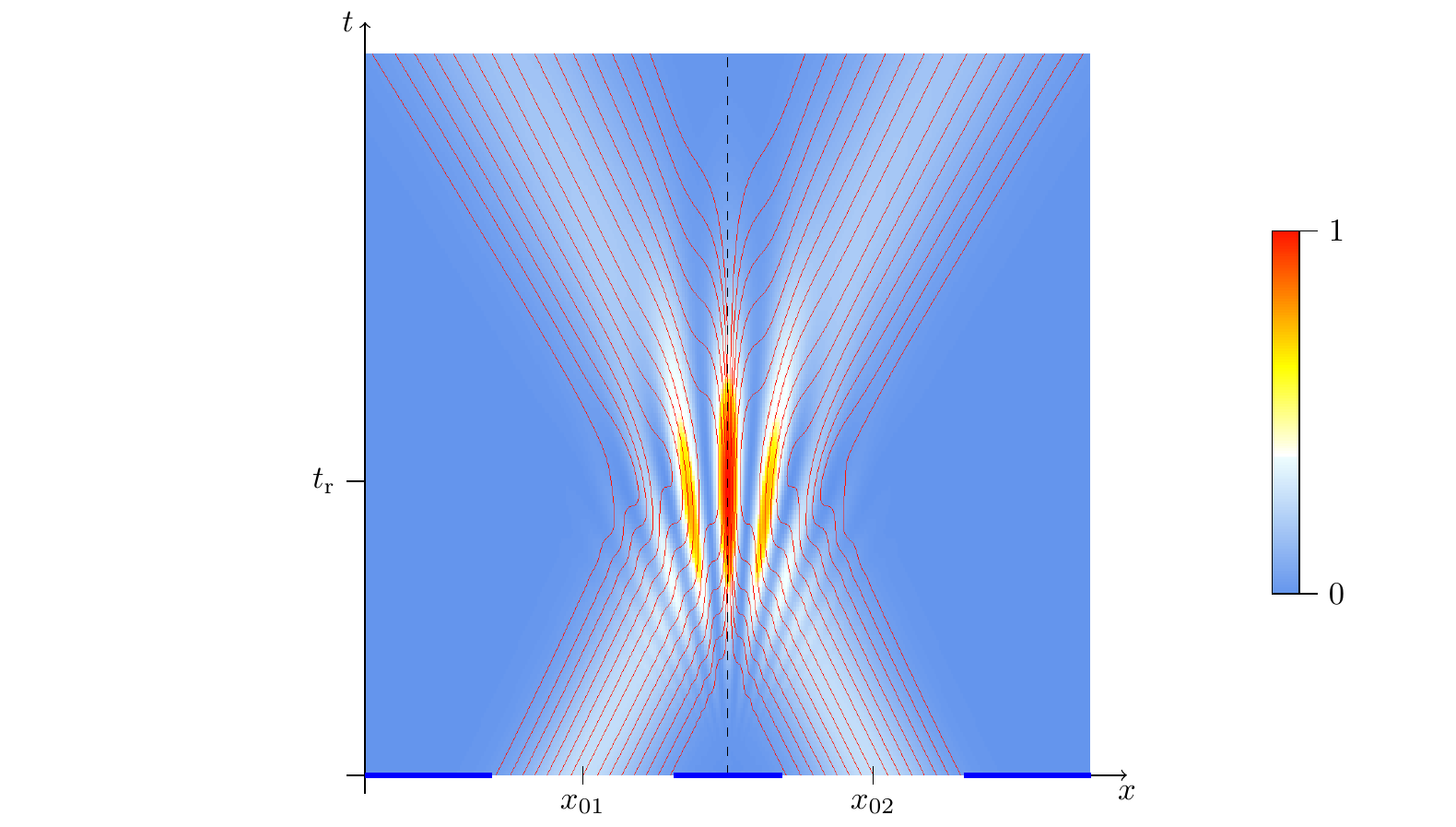}\caption{Classical computer simulation of the interference pattern in a double-slit
experiment; with $v_{x,1}=-v_{x,2}$.\label{fig:traj.2}}
\end{figure}
\begin{figure}[p]
\centering{}\includegraphics{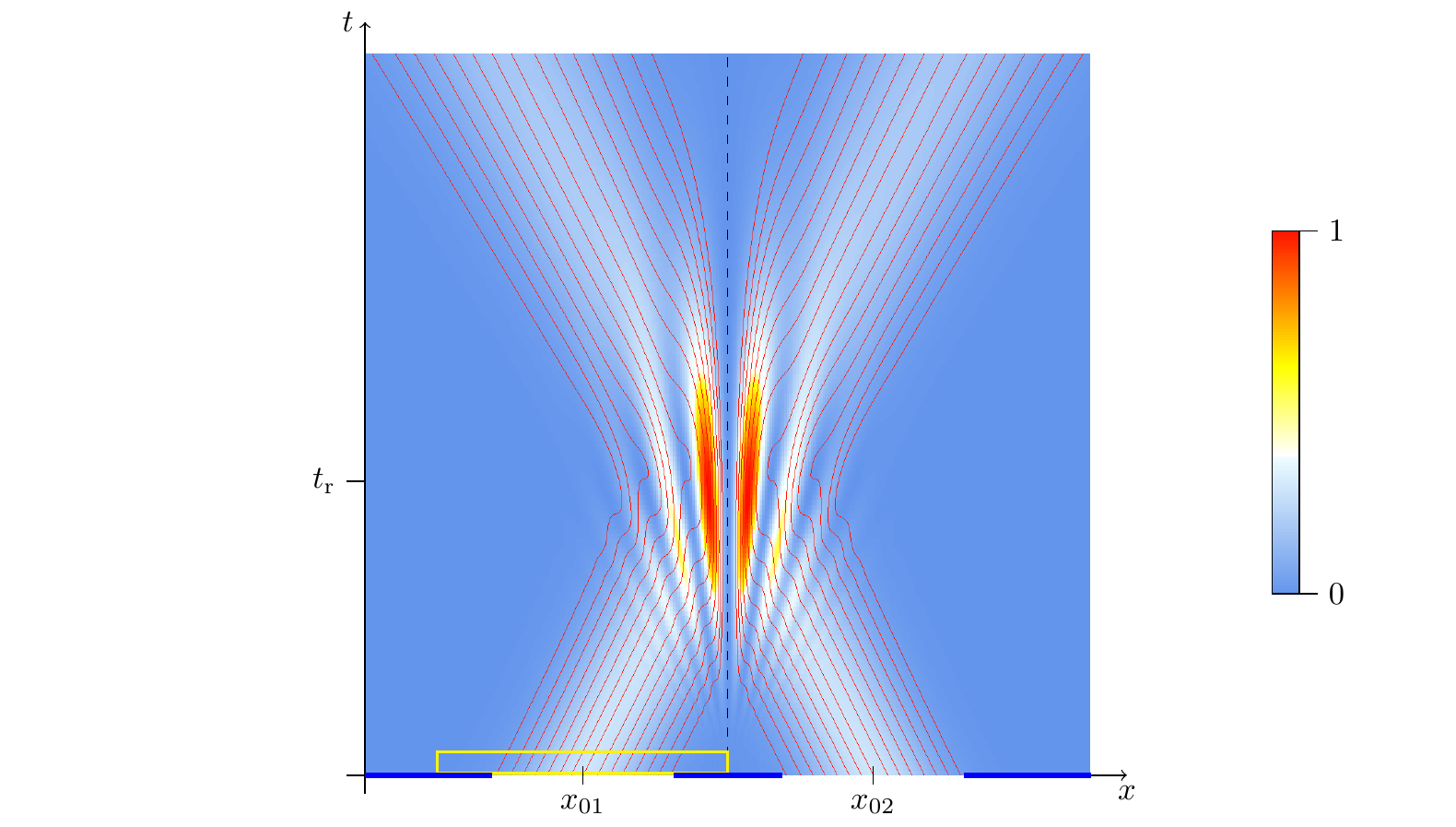}\caption{Same as Fig.~\ref{fig:traj.2}, with an additional phase $\Delta\varphi=\pi$
at slit~1.\label{fig:traj.3}}
\end{figure}
\begin{figure}[!ph]
\begin{singlespace}
\centering{}\includegraphics{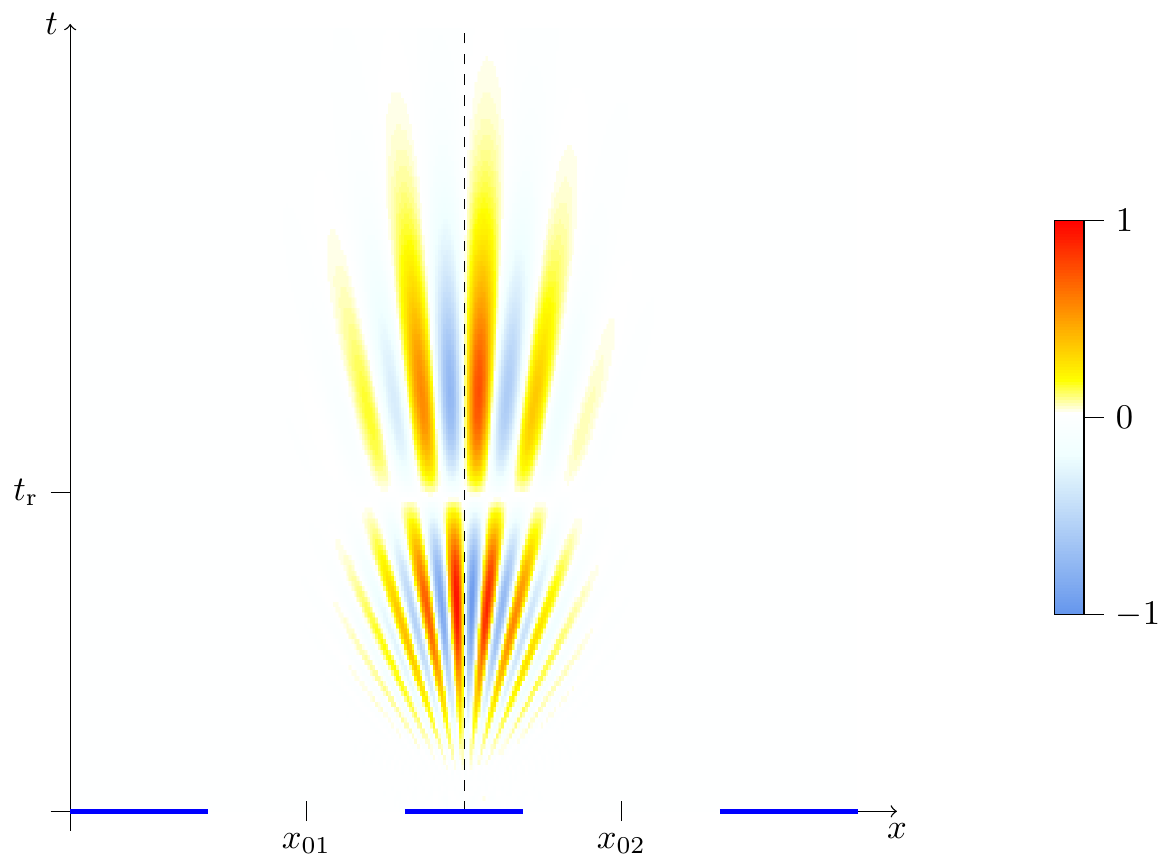}\caption{Classical computer simulation of the total average ``entangling current''
density in a double-slit experiment; same setup as in Fig. \ref{fig:traj.2},
with arbitrary normalization and $v_{x,1}=-v_{x,2}$.\label{fig:traj.4}}
\end{singlespace}
\end{figure}
\begin{figure}[!ph]
\begin{singlespace}
\centering{}\includegraphics{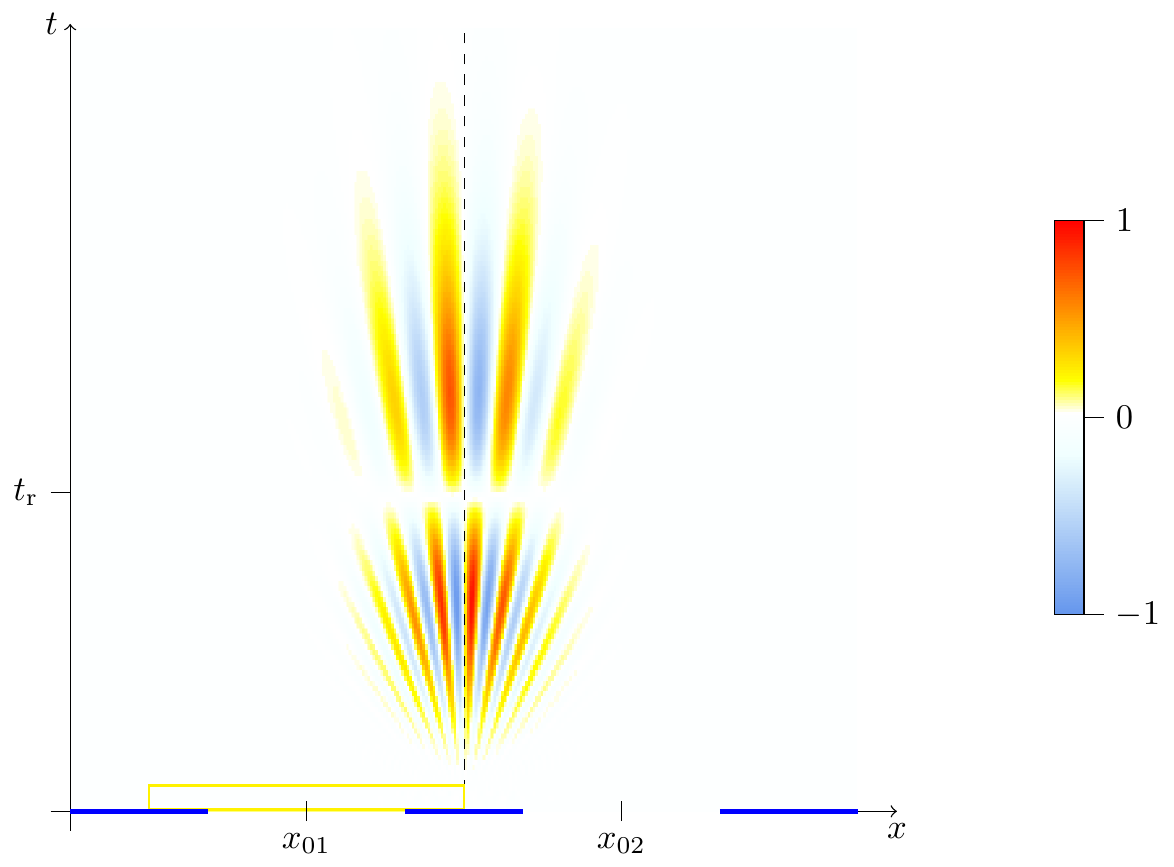}\caption{Same as Fig.~\ref{fig:traj.4}, with an additional phase shift of
$\Delta\varphi=\pi$ at slit~1 according to the setup of Fig.\ref{fig:traj.3}.
Comparing with Fig.~\ref{fig:traj.4}, one notes that the dramatic
shift from maxima to minima, and \emph{vice versa}, as observed in
the interference patterns of Fig.~\ref{fig:traj.2} and Fig.\ref{fig:traj.3},
respectively, is essentially caused by the changes in these entangling
currents.\label{fig:traj.5}}
\end{singlespace}
\end{figure}
where the phase difference 
\begin{equation}
\varphi_{12}=\varphi_{2}-\varphi_{1}=\frac{1}{\hbar}\left[m(v_{2}-v_{1})x+\frac{mu_{0}^{2}}{2}\left(\frac{(x-x_{02}-v_{2}t)^{2}-(x-x_{01}-v_{1}t)^{2}}{\sigma^{2}(t)}\right)t\right]\label{eq:vax.9}
\end{equation}
is characterized by the usual ``classical'' velocity difference
$v_{2}-v_{1}$and a kinetic energy term including a momentum fluctuation
$mu_{0}$, or the ``osmotic'' velocity $u_{0}$, respectively. Here,
we indicate the two slits at positions $x_{01}$ and $x_{02}$ and
we assume the same slit widths and hence the same initial standard
deviations $\sigma_{0}$. The graphical result of a classical computer
simulation of the interference pattern in a double-slit experiment,
including the average trajectories, with evolution from bottom to
top, is shown in Fig.~\ref{fig:traj.2}. The Gaussian wave packets
characterized by moderate spreading at the same standard deviations
$\sigma$ move towards each other with velocities $v_{x,1}=-v_{x,2}$.
One can observe a basic characteristic of the averaged particle trajectories,
which, only because of the averaging, are identical with the Bohmian
trajectories. To fully appreciate this characteristic, we remind the
reader of the severe criticism of Bohmian trajectories as put forward
by Scully and others (see \cite{Scully.1998bohm}, and references
therein). We can note that in our sub-quantum approach an explanation
of the obvious ``no crossing rule'' is straightforward and actually
a consequence of a detailed microscopic momentum conservation. In
Fig.~\ref{fig:traj.2} the maximum of the resulting distribution
is positioned along the central symmetry line in between the two slits.
\cite{Groessing.2012doubleslit}

The interference hyperbolas for the maxima characterize the regions
where the phase difference $\varphi_{12}=2n\pi$, and those with the
minima lie at $\varphi_{12}=(2n+1)\pi$, $n=0,\pm1,\pm2,\ldots$ Note
in particular the ``kinks'' of trajectories moving from the center-oriented
side of one relative maximum to cross over to join more central (relative)
maxima. In our classical explanation of double slit interference,
a detailed ``micro-causal'' account of the corresponding kinematics
is given. (For the details, see \cite{Groessing.2012doubleslit}.)
Since each Gaussian has its own phase distribution (\ref{eq:vax.9}),
we are free to add a phase shift $\Delta\varphi$ at one of the two
slits, say slit 1, which modifies $\varphi_{1}$ to
\begin{equation}
\varphi_{1}=\frac{S_{1}}{\hbar}+\Delta\varphi.\label{eq:vax.10}
\end{equation}
In Fig.~\ref{fig:traj.3}, we use the same double-slit arrangement
as in Fig.~\ref{fig:traj.2}, but now include a phase shifter affecting
slit~1, as sketched by the yellow rectangle on the left hand side.
The exemplary choice of $\Delta\varphi=\pi$ results in a shift of
the interference fringes. Comparing with Fig.~\ref{fig:traj.2},
we recognize now a minimum of the resulting distribution along the
central symmetry line.

Finally, we reconsider our classically obtained total average probability
current \cite{Grossing.2012quantum,Groessing.2012doubleslit}
\begin{equation}
J_{{\rm tot}}=P_{1}v_{1}+P_{2}v_{2}+\sqrt{P_{1}P_{2}}\left(v_{1}+v_{2}\right)\cos\varphi_{12}+\sqrt{P_{1}P_{2}}\left(u_{1}-u_{2}\right)\sin\varphi_{12}\,,\label{eq:vax.11}
\end{equation}
with osmotic velocities $u_{i}$ and convective velocities $v_{i}$
applied to both slits, $i=1$ and 2, and with the phases~(\ref{eq:vax.9})
and (\ref{eq:vax.10}).

The result of our computer simulation of Eq.~(\ref{eq:vax.11}) is
shown in Figs.~\ref{fig:traj.4} and \ref{fig:traj.5} corresponding
to the intensity distributions of Figs.~\ref{fig:traj.2} and \ref{fig:traj.3},
respectively. One recognizes the change of the maximum values of the
probability current along the central symmetry line in Fig.~\ref{fig:traj.5}
in comparison with those of Fig.~\ref{fig:traj.4}. Since the figures
display the one-dimensional case, the current flow is along the $x$--axis
only. Interestingly, at the time $t_{{\rm r}}$ of the reversal of
the trajectories, the current flow comes to a hold, and starts again
for times $t>t_{{\rm r}}$ with reversed signs. This can be understood
as a reversal of the relative flow of heat $Q_{2}-Q_{1}$ between
the two channels, since $u_{i}=-\frac{1}{2\omega m}\nabla Q_{i}$
\cite{Groessing.2009origin}, such that the last term of Eq.~(\ref{eq:vax.11})
reads as $\frac{1}{2\omega m}\sqrt{P_{1}P_{2}}\nabla(Q_{2}-Q_{1})\sin\varphi_{12}$.

The probability current $J_{{\rm tot}}$ in both figures essentially
only consists of the last term of Eq.~(\ref{eq:vax.11}), as the
velocities $v_{i}$ and the velocities $u_{i}$ typically differ by
many orders of magnitude. In other words, the probability current
$J_{{\rm tot}}$ is \textit{always} dominated by the ``quantum mechanical''
\textit{entangling} term of Eq.~(\ref{eq:vax.11}) which is connected
to the osmotic velocities, $u_{1}$ and $u_{2}$, and which implies
the existence of nonlocal correlations. As we have just seen, this
``entangling current'' can also be understood as describing the
``heat flow'' between the two channels. Note that, as opposed to
the average total probability current $J_{{\rm tot}}$, in the distribution
(\ref{eq:vax.8}) of the probability density $P_{{\rm tot}}$ alone,
just as in the correlated intensity (\ref{eq:vax.4}) of two-particle
interferometry, the entangling part is not explicitly visible.

The phenomenon of entanglement is thus possibly rooted in the existence
of the path excitation field, a version of which we already encountered
in the present paper with the usual double-slit interference. 

In other words, one can say that the entanglement characteristic for
two-particle interferometry is a natural consequence of the fact demonstrated
here, i.e., that already in single-particle interferometry one deals
with entangling currents, which generally are of a nonlocal nature.

\bibliographystyle{utphys}
\bibliography{../ains-reduced}

\end{document}